\documentclass[fleqn,usenatbib]{mnras}

\usepackage{graphicx}	
\usepackage{amsmath}	
\usepackage{amssymb}	
\usepackage{enumerate}
\usepackage{tablefootnote}
\usepackage[dvipsnames]{xcolor} 
\usepackage{newtxtext,newtxmath}

\title[Wide ECSNe]{Wide binary pulsars from electron-capture supernovae}

\author[Stevenson et al.]{
\newauthor  Simon Stevenson$^{1,2}$, 
Reinhold Willcox$^{2,3}$, 
Alejandro Vigna-G\'{o}mez$^{4}$ and
Floor Broekgaarden$^{5}$ \\
$^{1}$ Centre for Astrophysics and Supercomputing, Swinburne University of Technology, John St, Hawthorn, Victoria- 3122, Australia \\
$^{2}$ The ARC Centre of Excellence for Gravitational Wave Discovery,  OzGrav \\
$^{3}$ School of Physics and Astronomy, Monash University, Clayton, Vic. 3800, Australia \\
$^{4}$ DARK, Niels Bohr Institute, University of Copenhagen, Jagtvej 128, 2200, Copenhagen, Denmark \\
$^{5}$ Center for Astrophysics \textbar{} Harvard \& Smithsonian, 60 Garden Street, Cambridge, MA 02138, USA
}

\date{Accepted XXX. Received YYY; in original form ZZZ}

\pubyear{2021}

\begin{document}
\label{firstpage}
\pagerange{\pageref{firstpage}--\pageref{lastpage}}
\maketitle

\begin{abstract}
Neutron stars receive velocity kicks at birth in supernovae. 
Those formed in electron-capture supernovae from super asymptotic giant branch stars---the lowest mass stars to end their lives in supernovae---may receive significantly lower kicks than typical neutron stars.
Given that many massive stars are members of wide binaries, this suggests the existence of a population of low-mass ($1.25 < M_\mathrm{psr} / $M$_\odot < 1.3$), wide ($P_\mathrm{orb} \gtrsim 10^{4}$\,day), eccentric ($e \sim 0.7$), unrecycled ($P_\mathrm{spin} \sim 1$\,s) binary pulsars.
The formation rate of such binaries is sensitive to the mass range of (effectively) single stars leading to electron capture supernovae, the amount of mass lost prior to the supernova, and the magnitude of any natal kick imparted on the neutron star.
We estimate that one such binary pulsar should be observable in the Milky Way for every 10,000 isolated pulsars, assuming that the width of the mass range of single stars leading to electron-capture supernovae is $\lesssim 0.2$\,M$_\odot$, and that neutron stars formed in electron-capture supernovae receive typical kicks less than 10\,km s$^{-1}$.
We have searched the catalog of observed binary pulsars, but find no convincing candidates that could be formed through this channel, consistent with this low predicted rate.
Future observations with the Square Kilometre Array may detect this rare sub-class of binary pulsar and provide strong constraints on the properties of electron-capture supernovae and their progenitors.
\end{abstract}

\begin{keywords}
pulsars: general - transients: supernovae - supernovae: general - stars: neutron 
\end{keywords}



\section{Introduction}
\label{sec:intro}

Stars with initial masses $\lesssim 8$\,M$_\odot$ end their lives as white dwarfs, whilst more massive stars undergo core-collapse supernovae and form neutron stars or black holes \citep{Woosley:2002RvMP,Doherty:2017PASA}.
Super asymptotic giant branch \citep[SAGB;][]{GarciaBerroIben:1994ApJ,Doherty:2017PASA} stars\footnote{Recently \citet{OGrady:2020ApJ} reported the observation of $\sim$ 10 SAGB star candidates in the Magellanic Clouds.}
on the boundary between these two regimes are thought to result in electron-capture supernovae  \citep[ECSNe;][]{Miyaji:1980PASJ,Nomoto:1984ApJ,Nomoto:1987ApJ} leading to the formation of a neutron star.
The mass range of single stars expected to undergo ECSNe is narrow and roughly in the range 8--10\,M$_\odot$, although the exact mass range is very uncertain and depends on the details of the models \citep[e.g.,][]{Poelarends:2007ip,Doherty:2015MNRAS,Doherty:2017PASA,Jones:2016asr,Leung:2019phz}.
The mass range for ECSNe may be wider in interacting binaries \citep[e.g.,][]{Podsiadlowski:2003py,Doherty:2017PASA,Poelarends:2017dua,Siess:2018A&A}, which may result in the majority of ECSNe occurring in binary systems.
Whether ECSNe occur in single stars at all is an open question \citep[e.g.,][]{Willcox:2021kbg}.

ECSNe are expected to have low luminosities compared to other classes of supernovae, and be observed as type IIp or IIn supernovae \citep[e.g.,][]{Tominaga:2013ala,Moriya:2014uaa,Hiramatsu:2020obu}.
No observed supernovae have been conclusively associated with an ECSN, but SN2018zd may be the best candidate to date \citep{Hiramatsu:2020obu}.
Other candidates include the supernova that formed the Crab nebula/pulsar \citep[e.g.,][]{Nomoto:1982Nature,Smith:2013gya,Moriya:2014uaa}, though see \citet{Gessner:2018ekd}.


Numerical simulations succeed in realising ECSNe from first principles \citep[e.g.,][]{Kitaura:2005bt,Dessart:2006ApJ}.
Recently, \citet{Gessner:2018ekd} performed hydrodynamical simulations of ECSNe and found that the remnant neutron stars receive a kick of only a few km/s at most.
The small kicks arise from the rapid explosion, which does not allow time for substantial asymmetries to develop.
Kicks this small are much lower than the typical velocities of a few hundred km/s inferred for neutron stars from the proper motions of isolated Galactic pulsars \citep[e.g.,][]{LyneLorrimer:1994Nature, Hobbs:2005yx, Verbunt:2017zqi}.
For this reason, ECSNe play an important role in the formation of double neutron star binaries \citep[e.g.,][]{Vigna-Gomez:2018dza,Giacobbo:2018hze}, the retention of neutron stars in globular clusters \citep{Pfahl:2001df}, the population of wide neutron star high-mass X-ray binaries \citep{Pfahl:2002ApJHMXRB,Podsiadlowski:2003py,Knigge:2011Nature} and the mass distribution of neutron stars in binaries \citep[][]{Schwab:2010ApJ,Ozel:2012ax}.

Given that many massive stars are members of wide, non-interacting binaries \citep[e.g.,][]{MoeDiStefano:2017ApJS}, a population of neutron stars receiving (very) low kicks during ECSNe would lead to the prediction of wide, eccentric, unrecycled binary pulsars, which would not be predicted if ECSNe lead to large kicks (and thus disrupt wide binaries). 
Observations of wide binary pulsars (or the lack of these systems) could therefore be a powerful test of whether ECSNe can occur in effectively single stars. 
This could provide insights into ECSNe and their progenitors.

In this paper we develop a simple model for the formation of low-mass ($M_\mathrm{psr} \sim 1.3$\,M$_\odot$), wide ($P_\mathrm{orb} > 10^{4}$\,day), eccentric ($e > 0.5$), unrecycled ($P_\mathrm{spin} \sim 1$\,s) binary pulsars in the Galactic field formed by ECSNe. 
We describe our model in Section~\ref{sec:method}, and present results of a \textsc{Fiducial} model in Section~\ref{subsec:fiducial}.
We test the sensitivity of our predictions to model uncertainties in Section~\ref{subsec:variations}.
We estimate the number of wide binary pulsars observable in the Milky Way 
in Section~\ref{subsec:formation_rate}, finding that roughly one wide binary pulsar should be observable for every 10,000 isolated pulsars, assuming that the mass range of (effectively) single stars that lead to ECSNe is $<0.2$\,M$_\odot$ \citep{Willcox:2021kbg}.
In Section~\ref{sec:candidates} we search for candidate wide binary pulsars in the observed pulsar population.
We find no candidates that are well explained by our model, consistent with our estimated rates; the observation of these systems is hampered by the lack of sensitivity of pulsar observations to  orbital periods much longer than the observing baseline of a few decades.
We argue that a few (2--3) apparently isolated observed pulsars could actually be in wide binaries.
We conclude in Section~\ref{sec:conclusion}.


\section{Electron capture supernovae in a population of massive, wide binary stars}

\subsection{Method and fiducial assumptions}
\label{sec:method}



We model a population of wide, massive, non-interacting binary stars at solar metallicity, appropriate for pulsars formed recently in the Milky Way.
A large fraction of intermediate/massive stars are known to be part of binaries \citep[e.g.,][]{Sana:2012Science,MoeDiStefano:2017ApJS}. 
We are interested in the evolution of single $\sim 8$\,M$_\odot$ SAGB stars at solar metallicity, that may end their lives in ECSNe and produce observable pulsars in the Milky Way.
Given the uncertainties in the masses of SAGB stars which undergo ECSNe \citep{Doherty:2017PASA}, we assume\footnote{We have checked that the exact mass assumed does not impact our conclusions.} that all primaries (initially the most massive star in the binary) have a mass of 8\,M$_\odot$, as appropriate for solar metallicity SAGB stars.
We do not model binaries with primary stars with initial masses greater than 8\,M$_\odot$, as we assume that these will have already evolved and undergone core-collapse supernovae, receiving high kicks of order a few hundred km/s, resulting in the disruption of the binary \citep[see e.g.,][]{Vigna-Gomez:2018dza,2019A&A...624A..66R}. 

We sample the remaining properties of each binary from statistical distributions based on observations of massive, wide binaries. 
We denote our default set of assumptions our \textsc{Fiducial} model. 
We examine the predictions of this model in Section~\ref{subsec:fiducial}, and consider some alternate assumptions in Section~\ref{subsec:variations}.

We determine the mass of the secondary star by drawing the mass ratio of the binary $q = m_2 / m_1$ from a power-law distribution with a slope of $-2$ for mass ratios $q > 0.1$, favouring typical mass ratios of around $q = 0.2$, based on observations of massive, wide binaries \citep{Abt:1990ApJS,Sana:2014ApJS,Aldoretta:2015AJ,MoeDiStefano:2017ApJS}.

The binary orbital periods are drawn from a distribution which is flat in the log between $10^{4}$\,days (assumed to be the minimum orbital period of non-interacting binaries) and $10^{7}$\,days \citep{Opik:1924,Abt:1983ARA&A,Sana:2012Science,MoeDiStefano:2017ApJS}.
Wide binaries have low binding energies, and may be disrupted by flyby encounters with passing stars in the Galactic field \citep[e.g.,][]{Yabushita:1966MNRAS,Weinberg:1987ApJ}.
As discussed by \citet{Igoshev:2019cwq}, the typical lifetimes (before being disrupted this way) of the massive, wide binaries we consider here are longer than the $\sim 50$--$200$\,Myr the binary needs to survive to produce an observable binary pulsar (see Section~\ref{subsec:fiducial}).
Observations suggest that most massive wide binaries are eccentric \citep[e.g.,][]{MoeDiStefano:2017ApJS}.
In our \textsc{Fiducial} model we assume that the initial eccentricities of wide binaries are drawn from a \textsc{uniform} distribution between 0 and 1. 
We assume that the supernova occurs at a random time in the binary orbit.

The pre-supernova masses of SAGB stars are uncertain \citep[][]{Doherty:2017PASA}. 
For our \textsc{Fiducial} model, we assume that SAGB stars lose no mass during their pre-supernova evolution (this is the \textsc{No mass loss} model from Section~\ref{subsec:variations}).
This is approximately what occurs for stars in this mass range using the default mass-loss rate prescriptions in the binary population synthesis code COMPAS \citep{Stevenson:2017tfq,Vigna-Gomez:2018dza} that we use for estimating formation rates in Section~\ref{subsec:formation_rate}.
This implies that SAGB stars eject a large amount of mass during the supernova, which translates to a large mass-loss kick \citep{Blaauw:1961}.
Detailed SAGB models suggest that the envelope of these stars may be lost prior to the supernova through a brief phase with high wind mass-loss rates and thermal pulses \citep{Doherty:2017PASA}.
We therefore consider alternate assumptions regarding the pre-supernova mass loss of SAGB stars in Section~\ref{subsec:variations}.

We assume that all ECSNe give rise to neutron star natal kicks of the same magnitude.
In our \textsc{Fiducial} model, we assume that the magnitude of this kick is 10\,km/s, and we examine this in more detail in Section~\ref{subsec:variations}.
In reality, the natal kick will likely depend in detail on the properties of the progenitors.
However, to our knowledge, no such detailed prescription for ECSN kicks exists at present.
Our simple model is intended to allow us to easily understand the impact that natal kicks have on the binary\footnote{We expect that our results would be qualitatively similar if a narrow distribution of kick velocities was used}.
We make the standard assumption that the distribution of kick directions is isotropic \citep[e.g.,][]{Tauris:2017ApJ}. 
We determine the post-supernova orbital parameters following \citet{Kalogera:1996ApJ}.

ECSNe are formed from the lowest mass stars to go supernova, with core masses around the Chandrasekhar mass. 
The gravitational mass of the remnant neutron star is less than its baryonic mass by around 10\,\%, with the exact mass difference depending on the unknown neutron star equation of state.
We assume that the baryonic mass of all neutron stars formed through ECSNe is 1.36\,M$_\odot$ \citep[e.g.,][]{Takahashi:2013ena,Gessner:2018ekd}; for our assumed relation between the baryonic and remnant masses \citep{Timmes:1996ApJ}, this results in gravitational masses of 1.26\,M$_\odot$.


\subsection{Fiducial model predictions}
\label{subsec:fiducial}

To elucidate the main predictions of this model, we first consider the results of the \textsc{Fiducial} model described above. 
We examine the robustness of these results to  various assumptions in Section~\ref{subsec:variations}.

\begin{figure}
    \centering
    \includegraphics[width=\columnwidth]{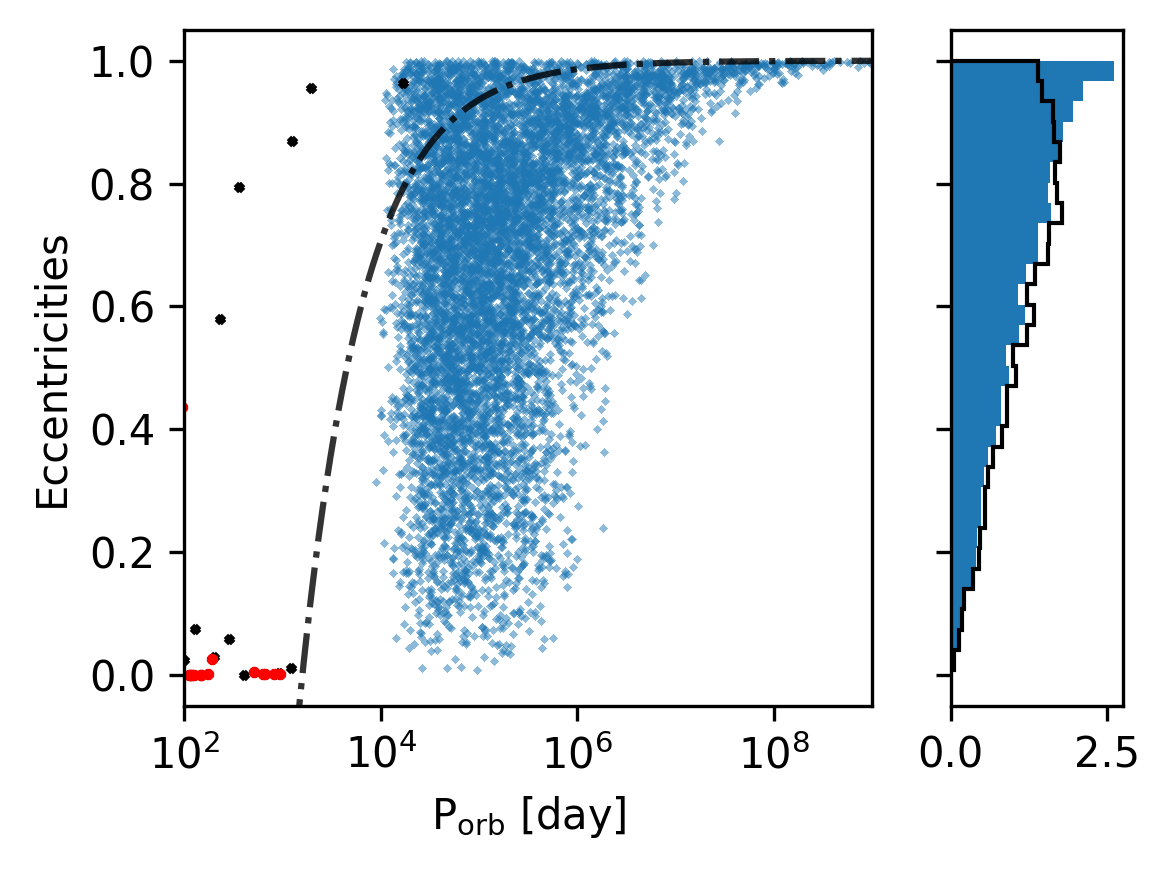}
    \caption{
    \textit{Left:} Orbital periods ($P_\mathrm{orb}$) and eccentricities of wide binaries after an ECSN in our \textsc{Fiducial} model (blue points; see Section~\ref{sec:method} for details). 
    The black dot-dashed line shows the eccentricities above which a binary would interact at periastron for a giant ($m_2 = 4$\,M$_\odot$, $1\,000$\,R$_\odot$) companion as a function of orbital period.
    The black points show the orbital periods and eccentricities of known binary pulsars \citep{2005AJ....129.1993M}. 
    Recycled pulsars are shown in red. 
    Uncertainties for the observed systems are typically much smaller than the size of the points.
    \textit{Right:} Normalised probability distribution of the eccentricities of wide binary pulsars. The solid black line shows the distribution after removing those binaries that would interact at periastron, as discussed in Section~\ref{subsec:fiducial}.
    }
    \label{fig:Porb_e_model}
\end{figure}

We show the orbital periods and eccentricities of wide binary pulsars following the ECSN in Figure~\ref{fig:Porb_e_model}.
We see that essentially all eccentricities are possible, although higher eccentricities are preferred; the median eccentricity in this model is around 0.7.
Around 10\% of binaries that remain bound are in high enough eccentricity orbits after the supernova that their periapsis separation would be equal to the radius of a low-mass giant companion ($R = 1000$\,R$_\odot$), potentially leading to mass transfer or a binary merger once the companion evolves. 
All other binaries do not interact, and therefore we expect the pulsar to be unrecycled and have typical properties of unrecycled pulsars, with a spin period $P_\mathrm{spin} \sim 0.1$--$1$\,s and a spin down rate of $10^{-17} < \dot{P} < 10^{-13}$ \citep[e.g.,][]{Boyles:2011ApJ}, unless ECSNe preferentially produce pulsars with particular rotational characteristics or luminosities. 
With the exception of the binaries with the highest eccentricities (which may result in mergers in any case), these binaries are sufficiently wide that gravitational radiation and tidal effects are not expected to significantly affect the binary orbit after the supernova on timescales when the pulsar is observable\footnote{The typical lifetime of an unrecycled pulsar is only 10--100\,Myr \citep[e.g.,][]{Lynch:2012ApJ,Chattopadhyay:2020lff}.}, and thus we neglect modelling these.
We therefore expect that the properties of observable systems should be very similar to their post-supernova orbital properties. 
We find that, because of the low kicks associated with ECSNe, these wide binary pulsars have low typical systemic velocities of $\lesssim 10$\,km/s.

The lifetime of an 8\,M$_\odot$ star is $\sim 40$\,Myr \citep{Pols:1998MNRAS,Hurley:2000MNRASSSE}.
Assuming a pulsar lifetime of 10\,Myr (100\,Myr; see above), this in turn implies that only companions more massive than 7\,M$_\odot$ (4\,M$_\odot$) can have evolved to massive ($M > 1$\,M$_\odot$) oxygen-neon white dwarfs whilst a pulsar is still visible. 
Lower mass stars will still be unevolved main-sequence stars.
In our \textsc{Fiducial} model, we find that only 2\,\% (10\,\%) of these systems should have massive oxygen-neon white dwarf companions with $M_\mathrm{WD} \sim 1$\,M$_\odot$, whilst the rest should have low-mass $M < 7$\,M$_\odot$ (4\,M$_\odot$) main-sequence companions.
The median companion mass ($\sim 2.8$\,M$_\odot$) arises from a competition between the initial distribution of binary mass ratios (which favours low-mass companions) and the supernova dynamics (binaries with a more massive companion are more likely to survive the supernova).

\subsection{Variations from fiducial assumptions}
\label{subsec:variations}

\begin{figure}
    \centering
    \includegraphics[width=\columnwidth]{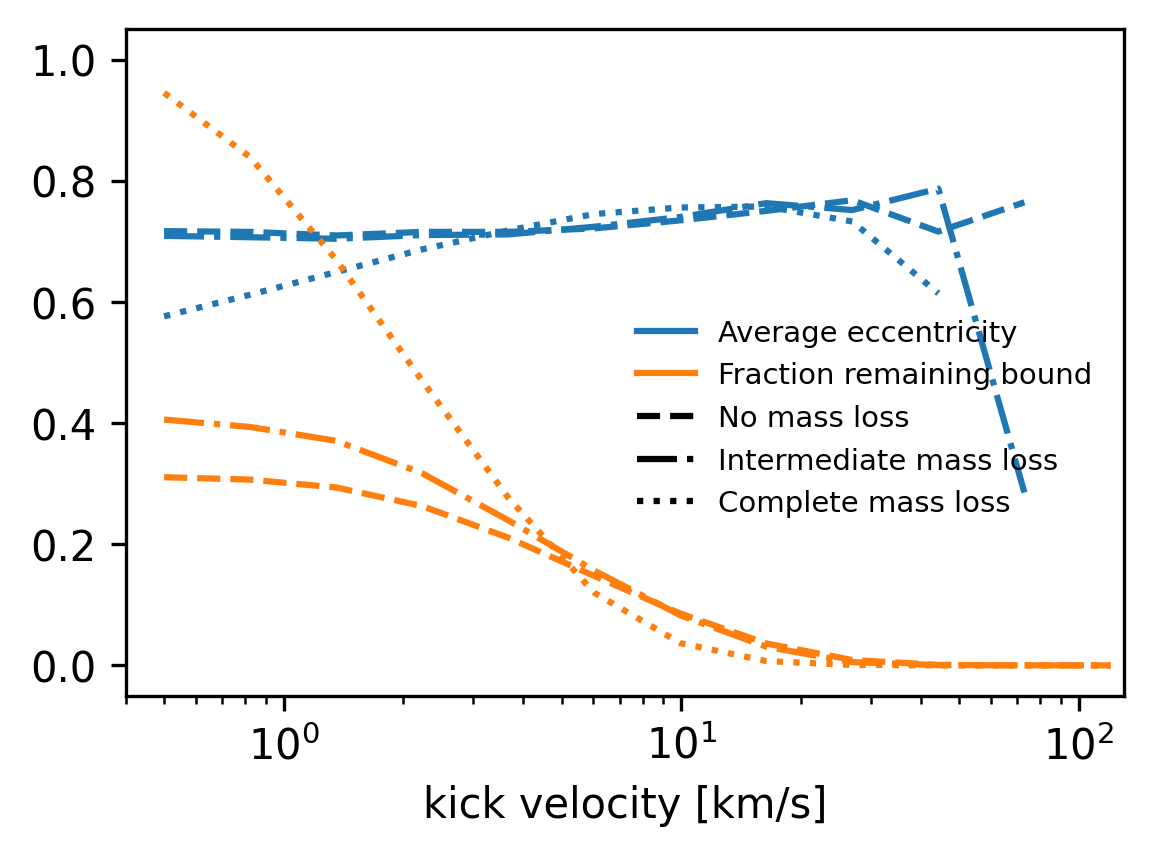}
    \caption{Fraction of wide binaries ($P_\mathrm{orb} > 10^{4}$\,day, $m_1 = 8$\,M$_\odot$) which remain bound after an ECSN (orange) and their average post-supernova eccentricity (blue), as a function of the kick velocity imparted to neutron stars in ECSNe.
    The different line styles show different assumptions about the amount of mass SAGB stars lose \textit{prior} to the supernova (see Section~\ref{subsec:variations}).
    The \textsc{No mass loss} model is shown with a dashed line, the \textsc{Intermediate mass loss} model is shown with a dot-dash line whilst the \textsc{Complete mass loss} model is shown with the dotted line.
    This figure shows results from the \textsc{Uniform} eccentricity model.
    }
    \label{fig:bound_fraction}
\end{figure}

Our model (described in Section~\ref{sec:method}) necessarily makes a number of simplifying assumptions. 
One of the key uncertainties is the pre-supernova (envelope) mass of SAGB star progenitors of ECSNe \citep{Poelarends:2007ip}.
Large pre-supernova masses will lead to large amounts of mass ejected during the supernova, leading to large mass loss kicks \citep{Blaauw:1961}.
In order to investigate the impact, we use three simple models.
The name of each model refers to the amount of mass loss the SAGB stars experience \textit{prior} to the ECSN.
In the \textsc{No mass loss} model we assume that the stars do not lose any mass through stellar winds prior to the supernova explosion. 
This is approximately what currently happens in binary population synthesis codes like \textsc{COMPAS} \citep{Stevenson:2017tfq,Stevenson:2019rcw,Vigna-Gomez:2018dza}.
Here, we do not intend this model to be realistic, but to demonstrate the extreme case.
At the other extreme is the \textsc{Complete mass loss model} where we assume that the pre-supernova mass is equal to the baryonic mass of the remnant neutron star, 1.36\,M$_\odot$ \citep{Vigna-Gomez:2018dza}. 
This model is motivated by detailed SAGB models that show that these stars have high mass-loss rates of order $10^{-4}$\,M$_\odot$\,yr$^{-1}$ during the final stages of their lives \citep[see][for a review]{Doherty:2017PASA}. 
However, we again expect that this model is too extreme.
In between these two extremes, in the \textsc{Intermediate mass loss} model we assume that the star experiences some mass loss and then ejects $M_\mathrm{ej} = 5$\,M$_\odot$ of mass during the supernova, based on models for the expected ejecta mass for ECSNe, along with that inferred from the Crab nebula  \citep{1997AJ....113..354F,Tominaga:2013ala,Hiramatsu:2020obu}.
For models that include pre-supernova mass loss we assume Jean's mode mass-loss, which leads to widening of the binary by a factor of a few.
We neglect wind mass transfer, which may be important even in wide binaries due to the slow wind speeds of SAGB stars \citep[e.g.,][]{Saladino:2018A&A,Hofner:2018A&ARv}.
We show the fraction of wide binaries that remain bound after the ECSN, and the average eccentricity of the remaining bound binaries for each of these models in Figure~\ref{fig:bound_fraction}.

The magnitude of kicks that neutron stars formed in ECSNe receive is another key uncertainty \citep{Gessner:2018ekd}.
In Figure~\ref{fig:bound_fraction} we vary the kicks from 0.5--120\,km/s.
In all cases we assume that all ECSNe lead to the same kick magnitude.
We find that for kicks less than 5\,km/s \citep{Gessner:2018ekd} at least 20\,\% of wide binaries would be expected to survive the supernova.
If ECSNe impart no kicks, around 30\% of binaries remain bound in the \textsc{No Mass Loss} model, whilst this fraction is close to 100\,\% for the \textsc{Complete mass loss} model.
Kicks $\gtrsim 10$\,km/s start to disrupt a large fraction ($> 90$\,\%) of wide binaries in all models, with essentially all binaries disrupted by kicks $>100$\,km/s (where the fraction remaining bound is $< 10^{-4}$). 
To summarise, this suggests that, given our current understanding of kicks from ECSNe \citep{Gessner:2018ekd}, a significant fraction of wide binaries should survive, and a sizable population of wide binary pulsars should exist.

We also show in Figure~\ref{fig:bound_fraction} how the median eccentricity of the wide binaries that remain bound varies with the kick velocity. 
All models lead to an average post-supernova eccentricity of $e \sim 0.7$ regardless of the mass-loss model or the magnitude of natal kicks.
We also find that our models with higher kicks than our \textsc{Fiducial} model result in slightly higher typical companion masses (since only stars with comparable mass companions survive the supernova).

Massive, wide binaries are observed to generally have eccentric orbits \citep{Malkov:2012A&A,MoeDiStefano:2017ApJS}.
In addition to our \textsc{Fiducial} model where the initial eccentricities of binaries are drawn from an \textsc{uniform} distribution, we have also tested the impact of assuming that binaries initially have \textsc{circular} orbits, or eccentricities drawn from a \textsc{thermal} distribution ($p(e) \propto e$).  
Eccentric binaries survive symmetric supernovae (i.e. with no natal kicks) more often than a circular binary of the same mass and orbital period due to the lower orbital velocity at, and greater time spent near, apoapsis.
The fraction of binaries which survive a symmetric supernova increases from around 20\% for the \textsc{circular} model to 50\% in the \textsc{thermal} model.
The initial eccentricity distribution is unimportant for natal kicks $\gtrsim 10$\,km/s.

Overall, with our 3 different assumptions regarding pre-supernova mass loss from SAGB stars, and 3 eccentricity distributions for massive binaries, we have simulated 9 different models with different assumptions. For each model we used 12 different values for the kick velocity of neutron stars in the range 0.5--120\,km/s, resulting in a total of 108 models.


\subsection{Estimate of the formation rate of wide binary pulsars}
\label{subsec:formation_rate}

ECSNe are rare events, constituting a few percent of the core-collapse supernova rate \citep{Poelarends:2007ip,Doherty:2017PASA,Jones:2018ule,Hiramatsu:2020obu}. 
The rate of ECSNe in wide binaries is sensitive to the mass range of (effectively) single stars that lead to ECSNe. 
\citet{Willcox:2021kbg} recently showed that the width of this mass range cannot be greater than 0.2\,M$_\odot$ in order to not overproduce low velocity pulsars compared to observations \citep[e.g.,][]{Verbunt:2017zqi,Igoshev:2020lif}\footnote{This also agrees with recent constraints on the width of the ECSN mass range from s-process element abundances in ultra faint dwarf galaxies \citep{Hirai:2019cpy,2021MNRAS.505.3755T}}.
Using the population synthesis models from \citet{Willcox:2021kbg}, we have calculated the formation rate of wide binaries in which the primary star is expected to undergo an ECSN i.e., the fraction of all binaries that are born in the parameter space described in Section~\ref{sec:method}.
We compare this to the number of isolated pulsars (which we assume have similar radio lifetimes, luminosities, beaming fractions etc) produced in the same model.
We find that one of these wide binaries is formed for every 100 isolated pulsars formed in our population synthesis models. 
A significant fraction of wide binaries are disrupted by the supernova in all of our models (cf. Figure~\ref{fig:bound_fraction}), with typically $\sim 10$\% surviving the supernova.
We therefore find one wide binary that remains bound following the ECSN (hereafter wide binary pulsars) is formed for every $\sim 1000$ isolated pulsars, leading to the possibility that $\sim3$ of the $\sim 3000$ observed isolated pulsars may have wide binary companions.
Wide binary pulsars may be misclassified as isolated pulsars due to their long orbital periods compared to the typical durations that pulsars have been observed (decades at most). 
Here we assume that binaries with orbital periods greater than 100 years would be misclassified as isolated pulsars \citep{Willcox:2021kbg}. 
In our models, we find that $\sim$ 1--10\% of these wide binary pulsars have orbital periods shorter than 100\,yr, with larger kicks generating a higher fraction.
Putting this all together, we estimate that one wide binary pulsar formed through an ECSN will be observed for every 10,000 isolated pulsars.
This is consistent with the current lack of observations of such systems (see Section~\ref{sec:candidates}), but raises the possibility of observing these systems with future pulsar surveys such as the Square Kilometre Array (SKA) that will expand the number of known pulsars to $> 10,000$ \citep[e.g.,][]{Keane:2014vja}.
Given that ECSNe make up only a small fraction of all supernovae, it is also worth considering whether a small fraction of neutron stars formed from more massive stars in more common core-collapse supernovae can also produce wide binary pulsars.
According to our models, for typical kicks greater than 100\,km/s \citep{Hobbs:2005yx,Verbunt:2017zqi}, less than 1 in $10^{4}$ binaries survive the supernova (cf. Figure~\ref{fig:bound_fraction}). This indicates that the dominant formation channel for wide binary pulsars will be through neutron stars formed with low kicks in ECSNe.
We emphasise that the rate estimates above are sensitive to several uncertain model assumptions (cf. Section~\ref{subsec:variations}).


\section{Search for candidate binary pulsars formed by electron-capture supernovae}
\label{sec:candidates}

Despite the low expected observation rate of wide binary pulsars (Section~\ref{subsec:formation_rate}), we have examined the pulsar catalogue to see if any known binary pulsars are consistent with having formed in this way. 
We do not find any compelling candidates, in agreement with our estimated formation rates.
We show the orbital periods and eccentricities of observed wide ($P_\mathrm{orb} > 100$\,day) binary pulsars in Figure~\ref{fig:Porb_e_model}, taken from version 1.65 of the ATNF catalogue \citep{2005AJ....129.1993M}, accessed using \textsc{PSRQPY} \citep{Pitkin:2018JOSS}.
The widest observed binary pulsar PSR J1024-0719 (not shown in Figure~\ref{fig:Porb_e_model} due to the large uncertainties in its parameters) has an orbital period of 2\,000--20\,000\,yr \citep{Bassa:2016fiy,Kaplan:2016ymq}.
PSR J1024-0719 is a 5\,ms pulsar \citep{Bailes:1997ApJ}, and thus has likely been recycled; its formation likely involves stellar dynamics (and hence it is not formed through the channel we propose here), and it may originate from a triple or a globular cluster \citep{Bassa:2016fiy,Kaplan:2016ymq}.
Other than PSR J1024-0719, the upper limit of orbital periods of known binary pulsars is  around $10^{4}$\,day (cf. Figure~\ref{fig:Porb_e_model}, see also \citealp{Igoshev:2019cwq}), with a preference for high eccentricities.

Many of the known wide binary pulsars are consistent with formation channels involving episodes of prior mass transfer, unlike the formation channel we consider here.
For example, PSR J0823+0159 \citep[B0820+02;][]{Manchester:1980ApJL,Hobbs:2004MNRAS,Xue:2017hgi} is a 0.8\,s pulsar in a wide (>1200\,day) binary with a low eccentricity ($e=0.01$) and a $0.6$\,M$_\odot$ white dwarf companion \citep{Kulkarni:1986ApJ,Koester:1992ApJ,Koester:2000A&A}. 
\citet{Tauris:2012MNRAS} propose that PSR B0820+02 formed from a wide low-mass X-ray binary.
In addition, several candidates have massive B/Be star companions; these are not compatible with our model since it predicts low-mass main-sequence companions.
One example of such a system is PSR J1740-3052 \citep{Stairs:2001MNRAS}, a young 570\,ms pulsar in a 230\,day, $e=0.57$ binary orbit with a massive ($M > 11$\,M$_\odot$) B star companion \citep{Bassa:2011MNRAS}.
Similarly, PSR J0045-27319 is a young binary pulsar with a 0.926\,s spin period, in a 54 day orbit with an eccentricity of $e = 0.8$ whose companion is also a B star of $\sim 10$\,M$_\odot$ \citep{Kaspi:1994ApJ,Bell:1995hr}.
PSR J1638-4725 \citep{Lorimer:2006qs} is a binary pulsar with an orbital period of 1941\,day and a high eccentricity $e > 0.95$. 
It has a companion mass of $\sim 8$\,M$_\odot$, which is likely again too massive to be explained by our model.
PSR B1259-63 (PSR J1302-6350, \citealp{Johnston:1992ApJ,Johnston:1994MNRAS}) is a  47\,ms pulsar with an orbital period of $1237$\,days, an eccentricity of $\sim 0.9$. 
PSR B1259-63 again has a massive ($M \sim 10$\,M$_\odot$) Be star companion, SS 2883. 
In this case both the spin period of the pulsar (indicating recycling) and the companion mass make it incompatible with our model.
PSR J2032+4127 is a 143\,ms pulsar in a wide (20--30\,yr) binary with a highly eccentric ($e > 0.8$) orbit \citep{Lyne:2015MNRAS}.
It too has a massive Be star companion ($\sim 15$\,M$_\odot$).
Another binary pulsar of interest is the recently discovered PSR J1954+2529 \citep{Parent:2021tbn}, which is a 0.93\,s non-recycled pulsar in a wide (82.7\,day), eccentric ($e=0.114$) orbit with a low-mass companion. 
The relatively close orbit of this binary (compared to the systems we have focused on in this paper) again suggests that this binary may have experienced a phase of mass transfer prior to the formation of the pulsar.
Since the primary has already formed a neutron star in these systems, it seems likely that they have undergone binary interactions in which some mass from the primary was transferred to the secondary prior to the supernova, particularly in the binaries hosting Be stars \citep[e.g.,][]{Vinciguerra:2020sdg}.

There is also an observed population of wide, unrecycled (young) binary pulsars known in globular clusters, which have been associated with ECSNe \citep{Lyne:1996ApJ,Boyles:2011ApJ,Lynch:2012ApJ}. 
However, the formation of these binaries likely involved dynamical interactions, rather than the isolated binary evolution channel we discuss here, so we do not examine them further here.

In conclusion, we do not find any observed binary pulsars that are well explained by our model.
This is likely either due to radio pulsar observations not being sensitive to orbital periods significantly greater than the observing duration, leading to wide binary pulsars being misclassified as isolated pulsars \citep{Bassa:2016fiy,Kaplan:2016ymq}, or because wide binary pulsars do not exist, either because the mass range of single SAGB stars leading to ECSNe is very narrow \citep[cf.][]{Willcox:2021kbg}, or because ECSNe lead to natal kicks $\gtrsim 10$\,km/s (cf. Figure~\ref{fig:bound_fraction}). 
These lead to the observed pulsar population being too small to observe these rare wide binary pulsars. 
Future radio observations with MeerKAT \citep{Bailes:2020qai} and the SKA \citep[][]{Keane:2014vja} will expand the known pulsar population, hopefully observing these systems.
It may also be possible to use GAIA to observe wide, low-mass main-sequence star companions to young pulsars \citep[e.g.,][]{Igoshev:2019cwq,Antoniadis:2020gos}.


\section{Summary and conclusion}
\label{sec:conclusion}

Both theoretical and observational evidence point to neutron stars formed in ECSNe receiving low kicks at birth \citep{Pfahl:2002ApJHMXRB,Gessner:2018ekd}.
We argue that if ECSNe occur in a population of wide, non-interacting binaries, low kicks predicts the existence of low-mass ($1.2 < M_\mathrm{psr} / $M$_\odot < 1.3$), wide ($P_\mathrm{orb} > 10^{4}$\,day), eccentric ($e \sim 0.7$), unrecycled ($P_\mathrm{spin} \sim 1$\,s) binary pulsars in the Galactic field.
These binary pulsars typically have low-mass main-sequence companions (Section~\ref{subsec:fiducial}).
Our model shows that at least 20\,\% of wide binaries are expected to survive an ECSN (cf. Figure~\ref{fig:bound_fraction}), if natal kicks are $\lesssim 5$\,km/s.
The exact fraction depends on the pre-supernova masses of SAGB stars, and the natal kicks imparted to neutron stars (cf. Figure~\ref{fig:bound_fraction}, Section~\ref{subsec:variations}).

We searched the catalogue of observed binary radio pulsars for systems with characteristics matching those described above (Section~\ref{sec:candidates}). 
We did not find any candidates which are well explained by this model.
Using binary population synthesis models \citep{Willcox:2021kbg} we have estimated the formation rate of these wide binary pulsars, finding that roughly one binary should be observable for every 10,000 isolated pulsars.
This is consistent with the lack of detection of such a system in the current population of observed pulsars, but raises the possibility of observing them in the future with the SKA.
Our low predicted observed rate is a result of a narrow mass range of single star ECSN progenitors \citep{Willcox:2021kbg}, combined with the fact that pulsar observations are not sensitive to orbital periods longer than a few $10^4$\,day.
The observation of a wide binary pulsar could provide evidence that ECSNe can occur for effectively single stars.

Key uncertainties remain in the modelling of ECSNe. 
These uncertainties include the mass range of progenitors which are expected to undergo ECSNe, what their pre-supernova masses are (for single stars when their envelopes are not stripped through binary interactions) and what magnitude kicks neutron stars receive in ECSNe. 
If the mass range of (single) SAGB stars leading to ECSNe is narrow, they will be inherently rare (or possibly even non existent), whilst ECSNe kicks larger than predicted by recent models \citep{Gessner:2018ekd} would disrupt most wide binaries (cf. Figure~\ref{fig:bound_fraction}).
We have assumed that all ECSNe form neutron stars, whilst some may lead to thermonuclear explosions or white dwarf formation  \citep[e.g.,][]{Jones:2016asr,Jones:2018ule,Tauris:2019sho,Leung:2019phz}.
Our discussion has focused on ECSNe, but the lowest mass iron core-collapse supernovae may also produce neutron stars with low kicks \citep[e.g.,][]{Muller:2018utr,Stockinger:2020MNRAS}.

In this paper we have focused on the formation of wide binary pulsars from wide, non-interacting binaries, since these would provide a clean probe of SAGB star evolution and ECSNe without any complications from binary interactions.
ECSNe are likely more common in close, interacting binaries \citep[e.g.,][]{Vigna-Gomez:2018dza,Vinciguerra:2020sdg,Willcox:2021kbg}.
Since a sizeable fraction of massive stars are in triples \citep{MoeDiStefano:2017ApJS}, these may also provide important contributions to the formation of wide binary pulsars through distinct evolutionary channels \citep[see e.g.,][]{Bassa:2016fiy,Hamers:2019oeq}.
Improved theoretical modelling of SAGB stars will aid in solidifying the theoretical predictions for populations of binary pulsars, whilst observing a large population of binary pulsars will place ever greater observational constraints on the evolution of these stars.


\section*{Acknowledgements}

We thank Ryan Shannon, Stefan Oslowski, Adam Deller, Eric Thrane, Ilya Mandel, Debatri Chattopadhyay and Ryosuke Hirai for useful comments and discussions.
We also thank the referee for constructive queries and suggestions.
The authors are supported by the Australian Research Council Centre of Excellence for Gravitational Wave Discovery (OzGrav), through project number CE170100004. 
SS is supported by the Australian Research Council Discovery Early Career Research Award project number DE220100241.


\section*{Data availability}

The results of this work are available from the authors at reasonable request.
This paper made use of results from the binary population synthesis code COMPAS (v02.19.02) as presented in \citet{Willcox:2021kbg}. The latest version of COMPAS is publicly available at \url{www.github.com/TeamCOMPAS/COMPAS}.



\bibliographystyle{mnras}
\bibliography{bib} 


\bsp	
\label{lastpage}
\end{document}